ARTICLE

# Incoherent? No, Just Decoherent: How Quantum Many Worlds Emerge

Alexander Franklin

King's College London, UK
Email: alexander.r.franklin@kcl.ac.uk

## Abstract

The modern Everett interpretation of quantum mechanics describes an emergent multiverse. The goal of this paper is to provide a perspicuous characterisation of how the multiverse emerges making use of a recent account of (weak) ontological emergence. This will be cashed out with a case study that identifies decoherence as the mechanism for emergence. The greater metaphysical clarity enables the rebuttal of critiques due to Baker (2007) and Dawid and Thébault (2015) that cast the emergent multiverse ontology as incoherent; responses are also offered to challenges to the Everettian approach from Maudlin (2010) and Monton (2013).

## 1. Introduction

The Everett interpretation of quantum mechanics is, at least according to Wallace (2012), a theory of many emergent worlds.[1] But exactly how do these worlds emerge? And is there a vicious circularity within the Everettian package that renders the metaphysics incoherent? I first demonstrate that the emergence of quasi-classical worlds in Everettian quantum mechanics can be described by the same philosophical formalism as is used to characterise emergence in other non-quantum contexts; I then build on this to defuse accusations of incoherence.

The analysis of emergence for Everettian worlds is somewhat under-developed in the literature. While Wallace appeals to a Dennettian real patterns framework together with the notion of instantiation, more needs to be said about exactly how these various ideas come together. In this paper I fill this lacuna, building on an approach to emergence developed elsewhere. I go on to demonstrate how this works by appeal to a model of the effects of decoherence in a quantum system.

Yet Dawid and Thébault (2015) claim that the standard conceptualisation of emergence within the modern Everettian framework is incoherent. The decoherence framework for emergence relies on the derivation of the relative smallness of interference terms after interaction with an environment, and some further argument is needed to justify the claim that such small terms are therefore negligible. Dawid and Thébault build on Baker (2007), Kent (2010), and Zurek (2003) to argue that such justifications are beset by vicious circularities since probabilistic reasoning is assumed to be required to evidence the claim that observers emerge: if the justifications rely on the decision-theoretic strategy, then it seems that appeal to observers is involved in the derivation of the existence of observers in the first place.

---

[1] Note that Everett was uncomfortable with such terminology; see Barrett (2011).

This worry relies on the claim that our evidence for quantum mechanics is essentially probabilistic. By considering a case study in which the predictions are non-probabilistic I show that this argument can be undermined. I contend that the neglect of terms with relatively small amplitudes can be justified non-probabilistically; as such, the circularity can be blocked. One might object to this line of reasoning by noting that the Born rule is also employed in the derivation of non-probabilistic predictions. My response to this worry is that in contexts where interference is rife, the probabilistic interpretation of the (mod-squared) amplitudes is ruled out, and that the Born rule, in such contexts, takes the form of an averaging measure rather than a probability measure. I liken this strategy to measures employed in other instances of emergence.

I go on to respond to other objections to Everettian emergence due to Maudlin (2010) and Monton (2013). In both cases these authors fail to recognise just how generic the style of emergence-based reasoning is within modern science, and thus how much of that corpus would need to be rethought if their claims were to be accepted.

Note that I leave controversies over the role and interpretation of probabilities to another paper, though this is of course well-trod territory; it's worth noting that I share various philosophers' qualms over the decision-theoretic approach to probability—the goal here is to argue that no interpretation of probabilities is required to establish the case for emergence of quasiclassical worlds. This paper focuses on the justification of emergence within the modern approach to Everett and argues that this can be accounted for independently of any particular solution to the probability problems.

In section 2 I appeal to an account of emergence developed elsewhere, and in section 3 I demonstrate how this gives rise to the emergence of Everettian worlds. In section 4 I cash this out by appealing to the example of the orbit of Hyperion—following Habib et al. (1998) and Berry (2001)—in this case application of decoherence theory predicts a determinate, classically chaotic orbit for Saturn's potato-shaped moon. In section 5, pace Baker, and Dawid and Thébault, I argue that the negligibility of interference terms as a consequence of decoherence does not require a probabilistic justification. In section 6 I respond to other objections to emergence in Everett. In section 7 I conclude that the Everettian emergence framework is coherent and posits a respectable metaphysics notwithstanding unsettled questions about the nature of its probabilities.

## 2. Emergence

Ross (2000) introduces "rainforest realism" to account for the many-levelled/layered ontology advocated by Dennett. Ross (see also Ladyman et al. 2007) expresses this in terms of abstractions from lower levels and higher-level projectibility. While the modern Everettians often appeal to this metaphysical framework, the details in, for example, Wallace (2012) are left somewhat sketchy. My goal in this section is to set out an account developed elsewhere that builds on Ross's, and demonstrate that this can render the emergence of the quantum multiverse more perspicuous. An upshot of my analysis will be that it enables responses to critiques of Everettian appeals to emergence.

Wallace (e.g. 2012, 2013) relies on the notion of "instantiation" to spell out the concept of emergence:

> Given two theories A and B, and some subset D of the histories of A, we say that A instantiates B over domain D iff there is some (relatively simple) map $\rho$ from the possible histories of A to those of B such that if some history h in D satisfies the constraints of A, then $\rho(h)$ (approximately speaking) satisfies the constraints of B.... The instantiation concept is much easier to illustrate than to define cleanly.... In the Solar System, molecular quantum physics (ignoring the measurement problem) instantiates classical mechanics: specifically, it instantiates the theory of classical point particles moving under an inverse-square force between them. (Wallace 2012, 54)

There's something clearly right about Wallace's suggestion that, for example, the instantiation relation between the dynamics of classical mechanics and aspects of molecular quantum physics help explain the emergence of one from the other. But, instantiation isn't sufficient for emergence—and Wallace (2022) provides reasons to think that he agrees with this claim.

In general, when we say that behaviour or dynamics emerges we suppose that there is a combination of both dependence and independence: without any dependence at all we would have absolutely distinct systems, and in the absence of any independence we should think of both systems as the same. It seems that Wallace's appeal to instantiation can, at least for sufficiently mathematised examples, explicate the dependence between the theories, but will be inadequate to illustrate their relative independence.

Another way of thinking about this is in terms of emergent entities: instantiation may be flexible enough that it would count all sorts of confected entities as emergent. For example, classical mechanics with no upper restriction on velocity instantiates classical mechanics in the same domain with some arbitrary upper limit (say 1000 ms$^{-1}$) on velocities, but, assuming that the instantiation of histories underwrites emergence of entities, no-one should think that all the entities of the restricted version of classical mechanics are emergent from those entities of the unrestricted theory.

We might look to Ladyman et al. (2007) to to spell out the relevant sense of emergence, but their employment of information-theoretic terminology is liable to be read as rather more epistemic than appropriate for realism about a quantum multiverse. However, the following is meant to build on their approach.

In Franklin and Robertson (2021), Katie Robertson and I develop an account of ontological emergence that remedies these issues. We specify the independence criterion and therefore develop an account of emergence that is harder to satisfy than mere instantiation and that results in emergent entities that we have reason to regard as genuinely adding to the ontology. This advantage is shared with Ross (2000): we distinguish those kinds in the special sciences that are to be included in the ontology from those which are not. However, we avoid information-theoretic terminology and are explicitly consistent with theoretical reductionism.

The upshot is that ontological emergence is shown to be rather less cheap than instantiation might sometimes seem. It's not the case that just any possible relation between the more and less fundamental underwrites emergence. After setting out this account of emergence, I'll show how it applies to the Everettian model.

Franklin and Robertson (2021) maintains that entities or kinds emerge if and only if they are involved in dependencies that are novel and screen off lower-level details. Screening off is understood as the combination of unconditional relevance and conditional irrelevance (see Woodward 2021); thus, our screening-off criterion entails that the higher-level dependencies are both dependent and independent of the lower level.

Suppose that $A$ is the height of the time-$t_2$ bounce of a bouncy ball, and $B$ is the height of the time-$t_1$ bounce of the ball, and the lower-level details (*LLD*) correspond to the configuration of the particles that constitute the ball at $t_1$. Then we can say that *LLD* is unconditionally relevant to $A$. That is, in the absence of any further conditions, the configuration of the particles at $t_1$ is relevant to the height of the bounce at $t_2$. Expressed more generally:

*Unconditional relevance:* conditional on a particular lower-level description, the probability of the macro-description $A$ obtaining increases: $P(A|LLD) > P(A)$. Under certain circumstances,[2] $P(A|LLD) = 1$.

The dynamical relation between the bounce heights at the two times encodes a macrodependency that screens off those lower-level details. In other words, the height of the bounce at $t_1$, and the dependency between the heights of the bounces at those two times described by the dynamics of bouncy balls, screens off the configuration of the particles at $t_1$. Conditional upon the dependency between the heights at the two times, the lower-level details are irrelevant.

*Conditional irrelevance (full generality):* $P(A/B \,\&\, LLD) \approx P(A/B) = x$, where $0 < x \leq 1$.

Conditional irrelevance amounts to a screening-off condition. We think it important that conditional irrelevance is approximate because there are circumstances in which the lower-level details of $A$ do not evolve into lower-level details subvening $B$. That is, screening off may not always be exact. A salient example of this is entropy-decreasing microstates in statistical mechanics. One might be tempted to respond to the observation that emergence is approximate with the rejection of ontological emergence altogether. My response is to note that, while metaphysical tastes vary, I do not know of any exact approach to emergence that would allow for the recovery of quantum particles, let alone gases and bouncy balls. In any case, one may view the project of this paper as conditional: if anything ontologically emerges, then the many quantum worlds and their distinct quasi-classical contents ought to count as emergent.

---

[2] If the microdynamics take all the members of the supervenience basis of $B$ to members of the supervenience basis of $A$.

The second condition for this account of emergence is novelty. In the context of mathematised sciences, the realisation of distinct dynamics is sufficient for novelty: the bouncy ball dynamics are novel because they have a distinct functional form from the dynamics of the molecules that constitute the ball. So an entity such as the bouncy ball is novel if it features in screening-off macrodependencies with distinct functional form from the corresponding microdependencies. In Franklin and Robertson (2021), we go into more detail regarding novelty for non-mathematised sciences, and relate this to novel causal powers. For the purposes of this paper, the current characterisation is sufficient; while one might concoct borderline cases of distinct functional form, all would likely agree that the dynamics of classical and quantum mechanics qualify as sufficiently distinct. This account of novelty is closely related to that developed in Knox (2016); Franklin and Knox (2018) as distinct dependencies will inevitably build on the variable changes emphasised by Knox.

In sum, entities or kinds weakly ontologically emerge if they participate in novel macrodependencies that screen off lower-level details. The novel dynamics of the bouncy balls and the state specified in terms of macroscopic variables and parameters screen off the microscopic configuration. Unconditional relevance is sufficient for instantiation, and Wallace's extra condition— that there is some relatively simple map which relates the macrohistory to the microhistory—will be satisfied if conditional irrelevance holds; if there were no relatively simple map then it would not, in general, be possible to identify macrodynamics or macrostates that screen off the microdetails.

With a clearer conception of emergence in hand, we can now discuss how worlds emerge in Everettian quantum mechanics.

## 3. Emergence and many worlds

According to the modern formulation of the Everett interpretation (see especially Saunders et al. 2010; Wallace 2012), the wavefunction evolves unitarily, and this evolution is uninterrupted by any dynamical collapses. Moreover, the wavefunction or the fundamental quantum state has the capacity to represent and describe all goings-on at relevant length scales, and so no additional hidden variables are required. Given the unitary dynamics, entanglement of macroscopic systems with those in microscopic superpositions in some basis will lead to macroscopic superpositions in that basis. The modern Everettian appeals to decoherence results to claim that, very rapidly, such macroscopic superpositions may be interpreted as effectively independent emergent worlds.

The question for this section concerns how the emergence of the worlds conforms to the characterisation of emergence given above. The answer to that question relies significantly on the use of decoherence as a mechanism for emergence—more specifically, decoherence underwrites the screening off of otherwise relevant details and is, thus, responsible for the screened-off states' evolving according to a novel (quasi-)classical dynamics.

The principal idea is that the quantum state has no preferred basis and consequently no preferred branching structure or splitting, while interference between branches in any given basis is dynamically significant. Insofar as there is no basis with respect to which individual branches evolve effectively independently from one another, there are no emergent worlds. Interference between putative branches guarantees the absence of such effectively independent dynamics. Non-constant phase relations are associated with spreading in configuration space (loss of overlap), which leads to interference being strongly suppressed. Given the numbers of particles involved in interaction with generic environments, this loss of local constant phase relations happens extremely quickly. In this paper I'll focus on environment-induced decoherence in which entanglement with external systems leads to non-constant phase relations, and induces suppression of interference; see Crull (2021) for a taxonomy of decoherence mechanisms.

This is further emphasised in Joos's contribution to Joos et al. (2013, 63–4):

> since scattering depends in an essential way on the position of the object, as in a microscope. Interference terms between different positions in the density matrix of the scattering centre are destroyed…["Destroyed"] means that certain interference terms are unobservable for "local" observations—so interference/phase information and superpositions between positions have very low amplitude/information is locally inaccessible.

While the system as a whole including the environment is still to be described as unitarily evolving according to the the Schrödinger equation, the effectively independent dynamics for any given branch lead to emergence. This conforms to the account above: the underlying unitary dynamics and interactions with other branches are screened off for the systems of interest by the branch-relative macrostate, giving rise to a novel macrodynamics.

In more detail, the screening-off criterion is satisfied as follows: the microphysical details are unconditionally relevant to the behaviour of any given macro-level system—the macroscopic behaviour is in principle derivable, and, as will be demonstrated shortly, model derivations are available for various systems. Focusing on the behaviour in a given Everettian branch, interactions with other branches are conditionally irrelevant if interference is suppressed as a result of decoherence—lack of interference guarantees effectively independent dynamical evolution on that branch. So, conditional on the macrostate in a given branch at some earlier time, but after a given branch splitting is dynamically preferred, subsequent evolution is screened off from much of the structure of the underlying quantum state. Note that, like other instances of emergence, the macrostate on which we conditionalise may not be straightforwardly expressible in microphysical terms—it's the change of variables to the macrostates that allows for the discarding of various microdetails.

Why do the microphysical details and other branches otherwise relevant to dynamical evolution not matter? As observed by Saunders (2021) the quantum contributions do matter in general—it's only in those cases where such contributions are conditionally irrelevant that we count the corresponding systems as emergent. The fact that there are such circumstances to be found is a consequence of the decoherence story just adumbrated.

The novelty criterion is satisfied insofar as we have classical dynamics describing a (quasi-)classical system in some circumstances. We can, in fact, describe our physical system differently as a consequence of the decoherence effects: rather than requiring the machinery of quantum mechanics, classical evolution is exhibited.

Examples of decoherence suppressing the interference terms and thereby leading to the conditional irrelevance of much of the structure of the underlying quantum state are well known; see, for example, Wallace (2012) for the Everettian story and Joos et al. (2013) for a detailed account of a variety of physical mechanisms for various kinds of decoherence. Insofar as interference is responsible for the deviation of quantum from classical behaviour, this then explains the novelty of the quasi-classical dynamics on each branch.

The physics of decoherence involves a great deal of theoretical and experimental complexity and there are many steps and assumptions which may be called into question. My goal in this paper is not to defend the general applicability of the physics of decoherence to real-world systems. In particular, one claim that I do not address is whether there are technical/mathematical reasons why interactions with environments for some systems will not lead to the selection of an effective basis and the suppression of interference. It's clear from the many models and many experiments performed that the physics of decoherence is successful and applies to many physical systems (see Schlosshauer 2019), but scaling these up to the ridiculously complex many-body systems we encounter in the real world is of course non-trivial. So this paper won't despatch worries of the form "the specific models for which decoherence has been demonstrated are special in some way or other," or that there are examples of physical systems which won't decohere on expected timescales.

Although I focus on environment-induced decoherence, it's worth noting that this isn't the only type. In fact, Halliwell (1998, 2010) and Wallace (2018) argue that conservation principles of various kinds can make it such that systems internally decohere. This means that, for example, if we have two distinct ground states of a system, and it is well described by a superposition of both of these, but there's some conservation rule that makes the transition between these states more difficult (or, in the limit, entirely ruled out) then each state is screened off from the other to a high degree of approximation. As a consequence, it's often appropriate to think of the dynamics of parts of such systems as emergent from the more fundamental system.[3]

Given how general the decoherence mechanisms are, the expectation is then that quasi-classical worlds emerge, where talk of the emergence of "worlds" is shorthand for the emergence of many interacting systems and entities, and their being screened off from other interacting systems. That is, we may identify within our physics effectively dynamically independent macroscopic physical systems that are described by distinct dynamics.

## 4. Case study

In order to demonstrate the sense in which we have screening off and novel dynamics, and thus emergent classical entities according to our account of emergence, it's best to use a case study. In Habib et al. (1998) the authors set out a model for environment-induced decoherence and extrapolate this to the case of Hyperion, discussed below.

---

[3] Vanessa Seifert and I consider an example of this kind of emergence for molecules in Franklin and Seifert (2020).

The model is based on the Hamiltonian in (1). This is taken from Lin and Ballentine (1990) where it's used to model a driven anharmonic oscillator and demonstrate how quantum effects are seen in systems which would be classically chaotic.

$$H = p^2/2m + Bx^4 - Ax^2 + \Lambda x \cos(\omega t). \tag{1}$$

Habib et al. use computational techniques to model the evolution of the Wigner function $f_W$—this characterises the structure of the quantum state in phase space. Equation (2) describes the master equation for the evolution of this system under decoherence assuming the weak-coupling, high-temperature limit of quantum Brownian motion, in which limit the contribution from decoherence can be encapsulated by the diffusion term $D$; in Zurek and Paz (1994) $D = 2\gamma M k_B T$, where $T$ is the temperature of the environment, $M$ is the mass of the system, and $\gamma$ is the relaxation rate (exchange of energy with the environment)—for more details see (Joos et al. 2013, chapter 3).

$$\frac{\partial f_W}{\partial t} = -\frac{p}{m}\frac{\partial f_W}{\partial x} + \frac{\partial V}{\partial x}\frac{\partial f_W}{\partial p} + L_q f_W + D\frac{\partial^2 f_W}{\partial p^2}, \tag{2}$$

where $\partial V/\partial x = 4Bx^3 - 2Ax + \Lambda\cos(\omega t)$ and $L_q$ is the quantum contribution due to quantum interference:

$$L_q \equiv \sum_{n \geq 1} \frac{\hbar^{2n}(-1)^n}{2^{2n}(2n+1)!}\frac{\partial^{2n+1}V}{\partial x^{2n+1}}\frac{\partial^{2n+1}}{\partial p^{2n+1}} = -\hbar^2 Bx\frac{\partial^3}{\partial p^3}.$$

If quantum interference is turned off, the classical ($L_q = 0$) Fokker–Planck equation is recovered in (3):

$$\frac{\partial f_c}{\partial t} = -\frac{p}{m}\frac{\partial f_c}{\partial x} + \frac{\partial V}{\partial x}\frac{\partial f_c}{\partial p} + D\frac{\partial^2 f_c}{\partial p^2}. \tag{3}$$

Habib et al. note that, while an isolated quantum system and a classical system will agree for initial times, quantum interference will soon dominate and the classical chaotic behaviour will be overwhelmed: "We have therefore good evidence that in isolated chaotic systems, the quantum–classical correspondence defined at the level of expectation values is lost relatively quickly due to dynamically generated quantum interference" (Habib et al. 1998, 4363).

This is depicted in figure 1, where the results of their computer model are presented. We see that, after a short time, the bottom image corresponding to classically chaotic systems depicting a solution of equation (3) deviates significantly from the top image, a solution of equation (2) with decoherence turned off. The final term in equation (2), representing decoherence, makes all the difference, restoring approximately classical behaviour to the quantum system, as depicted in the middle image of figure 1.

In the presence of $\sim 10^5$ particles we are assured that "[d]ecoherence destroys the interference pattern in the Wigner function, while at the same time, noise smooths out the fine structure of the classical distribution in such a way that quantum and classical distributions and expectation values both converge to each other" (4363–4).

The diffusion term (the fourth term on the right-hand side of (2)) leads to decoherence because the chaotic dynamics engender spread of the quantum system in position space, and, thus, its narrowing in momentum space. For any system sufficiently localised in momentum space, the diffusion term generates a loss of coherence, which leads to the suppression of interference between each emergent, classically chaotic branch.

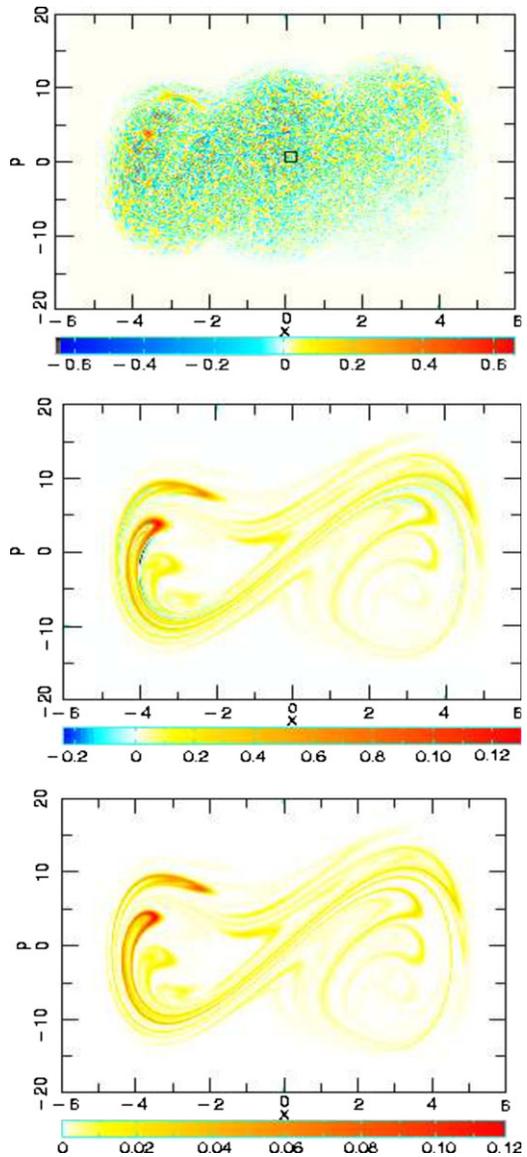

**Figure 1.** Top: Solution of (2) with time $t = 8T$, where $T$ is the period of the driving force, $D = 0$, quantum interference with no decoherence. Middle: Solution of (2) with $t = 8T$, $D = 0.025$, decoherence suppresses quantum interference. Bottom: Solution of (3), $t = 8T$, $D = 0.025$, classical distribution (Habib et al. 1998).

If their analysis is correct, then this gives us reasons to believe that classically chaotic systems emerge from quantum systems. This satisfies the account of emergence given above: first, the top diagram in figure 1 demonstrates the unconditional relevance of the quantum effects—interference modelled by $L_q$ in (2) is unconditionally relevant to the dynamics of the system; second, the middle diagram shows that decoherence can suppress such contributions, rendering the quantum contributions irrelevant conditionally on the presence of decoherence and the state of the system after times when decoherence effects dominate; third, the approximate agreement of the middle and bottom diagrams demonstrates that the emergent system is well modelled by classical dynamics.

As such, we have a classical system emerging from the lower-level quantum system. In terms of the Everett interpretation, following (Wallace 2012, chapter 3), decoherence has the effect of continually splitting the quantum system into classical emergent systems that exhibit classically chaotic dynamics, each of which is screened off from much of the structure of the underlying quantum state.

Now that we have good reason to think that decoherence leads to emergence in a model system, what justification is there for the claim that this is instantiated in the world? An intriguing series of claims initiated by Zurek (1998) and Berry (2001) suggest that we can think of Hyperion—the moon of Saturn—as well described by dynamics of this form.[4]

On the assumption that Hyperion is, fundamentally, described by quantum mechanics, the time taken for its classically and quantumly described orbits radically to differ in the absence of decoherence is less than 20 years. Notwithstanding the fundamentally quantum nature of the matter of which Hyperion is constituted, we observe it to have a classically chaotic orbit; this may then be taken to be a consequence of interaction with an external environment—Berry (2001) argues that just the photons from the sun lead to a decoherence time of $10^{-53}$ seconds (although Berry notes that the precise number should not be taken seriously; the "extreme smallness" of the number should be). As such, we may think of the observed classically chaotic orbit of Hyperion as observable evidence of the effects of decoherence in suppressing quantum interference.[5]

Classically chaotic Hyperion counts as emergent because much of the structure of the underlying quantum state is conditionally irrelevant to the future dynamics of each classically chaotic Hyperion. In macroscopic terms, what's screened off are the interference terms that would describe interactions with the Hyperions in other branches—rendering the other branches irrelevant to each branch's evolution. And the classically chaotic dynamics is not instantiated in the quantum system absent environment-induced decoherence. Much work is required, of course, to extend this analysis fully to evidence the claim that there are many emergent quantum worlds, but the hope is that this and the previous sections have made clear in which sense such worlds may be said to emerge, if they in fact do!

## 5. Incoherence?

Baker (2007), Dawid and Thébault (2015), and Zurek (2003) offer closely related critiques of the theoretical package advanced by modern Everettians, exemplified especially by Wallace (2012). The argument is that there is an incoherence or a vicious circularity inherent in the package that includes both the decision-theoretic—and thus observer-focused—justification of the application of probability in combination with a naturalised—and thus observer-independent—conception of emergence. In brief, my response is that no interpretation of probabilities is required in order to establish the case for emergence.

---

[4] Although this analysis has been called into question by Wiebe and Ballentine (2005), I agree with Schlosshauer (2008) that such objections rely on the ensemble interpretation, and that this interpretation is unacceptable since it cannot solve the measurement problem.

[5] Berry attributes emergence of the classical to the appeal to a singular limit; see Franklin and Knox (2018) for the argument that emergence needn't rely on such limits.

Dawid and Thébault (2015, 1566) start by asking a series of worthwhile questions, some of which I've addressed above:

> [i]n essence, "[d]ecoherence causes the universe to develop an emergent branching structure. The existence of this branching structure is a robust (albeit emergent) feature of reality" (Wallace 2012, 101). The key question, of course, must be: what do the terms "effective independence," "emergent," and "robust" mean here? On our analysis...these notions can only be of use to physical science if the sense in which they may be empirically grounded is made precise.

I agree that such concepts not only must be sharply defined, but that empirical grounding is also required. To foreshadow my response, I take it that consideration of cases like that of Hyperion go some way to providing such empirical grounding. However, Dawid and Thébault continue:

> It is difficult to see how such empirical grounding can be provided without assuming some version of the Born rule as a measure of probability of experimental outcomes. Thus, save for the provision of new, non-probabilistic empirical grounding, the proponent of the DWE [Deutsch–Wallace–Everett] scheme must, in deploying strategy b) [indirect justification for neglecting small amplitudes when these represent interference terms], still assume the Born rule at some level...
>
> **Z** Decoherence effects cannot be established without an independent prior derivation (or assumption) of the Born rule.

They argue that the justification of the emergence of branches and Everettian worlds presupposes some prior association of small amplitudes with small effects— otherwise one could not discard the interference terms and postulate multiple effectively independent outcomes. They suppose that the only way to associate small amplitudes with small effects is to regard the small-amplitude terms as improbable— and that, as such, an analysis of probability is assumed. My response is that we may reject principle **Z** and justify emergence more directly. That's because there are some predicted effects that are not probabilistic and that the screening off of interference is a dynamical rather than a probabilistic effect. Therefore, an interpretation of probability is not required in order for the emergence claims to be appropriately evidenced. In other words, I deny their claim that one must independently justify the association of small amplitudes with small effects—I claim that the account of emergence offered in the previous two sections is empirically grounded and did not need to assume such an association.

In order to defend these assertions it's best to re-express their argument in the terms developed above. They claim that we have no good basis to assert that the quantum effects and interaction with other branches are conditionally irrelevant, and, consequently, we aren't justified in postulating screening off or quasi-classical emergent dynamics.

Their worry becomes particularly pressing when it's noted that screening off is never absolute: the classical world emerges approximately—in this case emergence is based on discarding the heavily suppressed interference terms between the putatively emergent worlds. They note that empirical justification is required for the claim that the small amplitudes (coefficients of the interference terms) may be neglected—that small amplitudes correspond to, or represent, negligible contributions. If we had no such justification then the claim that the other branches are conditionally irrelevant would be ungrounded.

As Dawid and Thébault observe, a salient way within the literature to justify neglecting the small-amplitude contributions is based on the Born rule that links mod-squared amplitudes to probabilities. If the Born rule measure may be employed and justified via, for example, the decision-theoretic arguments that lead to the Deutsch–Wallace theorem, then we would be justified in assigning extremely small probabilities to such terms and thus discarding them. However, Dawid and Thébault claim that we may not employ the decision-theoretic arguments in the absence of distinct observers on well-defined branches, and there's no reason to think that there are such branches unless we can justify emergence by appeal to probabilities. Hence the circularity.

I agree that decision-theoretic arguments presuppose the existence of effectively independent, emergent observers; if this were the only available justification for the irrelevance of the low-amplitude terms then I grant the circularity. The Everettian approach would then be rightfully accused of pulling itself up by its bootstraps. I argue below that the existence of emergent branches can be established without any appeal to probabilities—as such, the interpretation of probabilities in this approach can be postponed to another occasion.

An essential assumption for the circularity claim is principle **Z**, and it's this that I wish to call into question. What if decoherence effects are established via the derivation of classical behaviours? That is, might decoherence and the emergence of the classical be empirically sensitive ontological claims that do not require or presuppose an "independent prior derivation of the Born rule?"

The thought is that we gain knowledge of decoherence effects both via specific experiments and more generally by the observation of the behaviour of bodies, such as Hyperion, whose dynamics would be different in the absence of decoherence. The specific predictions made—that Hyperion's orbit will be classically chaotic—do not require reference to probabilities and are not premised on any particular application of the Born rule qua probability measure. As such, the emergence of worlds in Everett can be established in a way that does not lead to a vicious circularity of any kind, and this is the precise sense in which emergence claims may be empirically grounded. No probabilistic claims were required in order to connect decoherence with the emergence of worlds.

One general question that this raises is to what degree our evidence for emergent theories floats free of the theory from which they are emerging. Given screening off and consequent effective autonomy we might be sceptical concerning the degree to which we may confirm a more fundamental theory by observing emergent phenomena—in this case, if Everettian quantum theory could be evidenced by the observation of classical phenomena. However, what Dawid and Thébault's critique calls into question is whether the fundamental account could, even in principle, give rise to the phenomena observed. Thus, observations coupled with a derivation of classically chaotic phenomena from the underlying theory ought to do much to settle the qualms of such critics.

I'm not arguing that, in general, fundamental theories are to be confirmed by the emergent theories to which they give rise, although some increase in the warrant for both theories will generically come about. Rather, I'm claiming that we may answer the sceptic who claims that an Everettian quantum theory *cannot* give rise to emergent worlds.

Baker's earlier paper expresses a similar worry in the form of a trilemma for interpreting the quantum state post decoherence. He notes that, if decoherence sent interference terms exactly to zero, the Everettian would be able to talk coherently of distinct branches and outcomes. However, the physics does not permit such a simple story. As a result the relevant options are *either* to regard the interference terms as distinct branches (*or* as parts of branches) in addition to those representing the observed effects and to disregard them because they are "improbable," *or* to disregard interference terms because the branches corresponding to the observed effects plus the interference terms are "close in Hilbert space to the non-interfering branches [where interference terms are exactly zero]" (Baker 2007, 163).

My response to both critiques is to acknowledge the circularity of the story as they present it. Dismissing interference terms because they are improbable would indeed require a prior justification of the Born rule. On the other hand, the story told in the previous section is not thus threatened by circularity, for two reasons. First, this project is less ambitious than the full Everettian project: at some stage, of course, one needs an account of the role of probability within the Everettian picture, but here I simply claim that one can account for emergence in Everett without providing an account of probability. So what I've done is to demonstrate that some classical behaviour emerges in some cases and that we have both modelling and observational evidence that attests to that. Second, by focusing on emergence and screening off as a dynamical phenomenon—that the interference terms don't make a dynamical difference to the evolution of the classically chaotic system—we can justify their irrelevance without relying on the claim that they are improbable.[6]

An immediate objection to these claims will occur to many readers. Even if the predictions used to confirm the theoretical package are not probabilistic, and thus do not require an independent, prior justification of the Born rule, the determination of the effects of interaction with an external environment, and the prediction of the effects of decoherence in any given case, will require appeal to averages, expectation values, and other implicit uses of the Born rule. In this case the circularity may not be avoided because the derivation itself requires the Born rule and, if the only justification of this is via decision-theoretic arguments, little progress will have been made.

---

[6] This is closest in spirit to $B_3$ in Baker (2007).

The analysis in Habib et al. (1998) and the more detailed derivation in Zurek (1998) appeal to expectation values and to various approximations that may be seen to rely on assumptions that there is a close relationship between small amplitudes and negligibility of dynamical contributions. Does this, then, implicitly appeal to a probabilistic analysis? Even if some justification for probabilities that did not rely on decision-theoretic reasoning could be provided, the controversy associated with probability in the Everett interpretation is such that this would threaten the cogency of the emergence claims.

My response to such worries is that multiple uses of the Born rule, or Born rule measure, must be distinguished. That's because a probabilistic interpretation of the mod-squared amplitudes is inapplicable before decoherence has occurred; see, for example, Wallace (2014, 207). In the presence of interference, amplitudes may cancel each other out—interpreting amplitudes in such contexts probabilistically will not do. It is only when interference is sufficiently suppressed that mod-squared amplitudes approximately conform to the probability axioms: any attempt to interpret mod-squared amplitudes as probabilities in the presence of interference will be empirically undermined as can be seen, for example, if one imposes a probabilistic interpretation half way between the double slit and the screen in the double slit experiment.

Therefore, at least in some of the contexts where the Born rule measure is applied and expectation values are discussed these are not to be given a probabilistic interpretation, and they certainly shouldn't be associated with a decision-theoretic justification! Rather, we should think of the relation between small amplitudes and irrelevance as a dynamical phenomenon. The relative magnitude of the amplitudes encodes the dynamical contribution of each term. So Baker, and Dawid and Thébault, were wrong to think that we must justify discarding interference terms by stating that they are improbable. Rather, they are discarded because, in some contexts (those where we find emergence), they hardly make a difference to the dynamical evolution of the on-diagonal terms. The claim that classical behaviour emerges is not a priori; the fact that we have models where the evolution of some behaviour is screened off from the evolution of the rest of the branch is a substantive discovery, and one that justifies the Everettian emergence claims.

This is in no way unusual: in every context of emergence and inter-theoretic relations some or other measure is employed to relate the many lower-level variables to the fewer higher-level variables—a process referred to by Wilson (2006) as "variable reduction."

Consider the bouncy ball discussed above. The centre of mass variable—the result of a measure that assigns a weight to each equal-mass contribution to the total mass in proportion to its distance from the centre—is dynamically preferred, and is used to construct the higher-level description. One can gain justification for this claim by inspecting the dynamics and various forces, along with salient approximations; for example, if the gravitational field appreciably varies over the height of the object this measure would need to be modified. But the justification for the appeal to this measure is empirical—it works and is empirically grounded by the testing of the predictions made on that basis.

Likewise, we look to Hyperion and similar such examples to provide justification of the cogency and coherence of this instance of emergence and the measures employed in the derivation of emergent behaviour. Of course, such cases alone would not be sufficient to provide an empirical grounding for Everettian quantum mechanics in its entirety. In fact, I think that something along the lines of the empirical robustness and triangulation discussed in Wimsatt (2007) and Evans and Thébault (2020) is the right way to conceive of our evidence for this approach.

When faced with instances of inter-theoretic relations, the predictions of our theories may be confirmed by various distinct experiments and observations. So, for example, in medical science the high-level observation of correlations between, say, smoking and lung cancer, and the low-level mechanistic understanding of the effects of tar on the cells of the lung, mutually reinforce and lead to an understanding of individual claims as especially robust. While a full discourse on the epistemology of science is beyond the scope of this paper, it should be clear that Everettian quantum mechanics as a research programme gains evidence from the macroscopic observations and models discussed above, from the many empirical successes of quantum theory, and, in particular, from experimental work on decoherence, referenced and discussed in Schlosshauer (2019). In cases where the emergent predictions and observations are not probabilistic, no mention of the Deutsch–Wallace theorem is required.

I should note that this response is very much in line with Wallace's brief reply to Baker (2007) and Kent (2010)—he notes that we don't do anything special in the Everett context that isn't done in other contexts of emergence, but that more could be said about emergence itself. "I'll concede that there's plenty of work to be done here by philosophers of science with an interest in emergence" (Wallace 2012, 254). My hope is that in this paper I've made some progress in this respect.

Dawid and Thébault's contention is that emergence claims cannot be justified without adducing some kind of relationship between probabilities and amplitudes. I argue that the observation of Hyperion's orbit justifies the dynamical relations used in the derivation of the emergent behaviour. Any measure that preserves some kind of connection between amplitude and dynamical salience will, thus, be sufficient to underwrite the emergence claims, without appeal to probabilities.

Up to this point I've sought to establish that applications of a Born rule measure, or, more generally, of the connection between small amplitudes and dynamical irrelevance, may be justified in much the same way as other measures in other parts of science, without requiring reference to probabilities.

However, mod-squared amplitudes *are* sometimes appropriately interpreted as probabilities. So while I've shown that there's evidence of emergence that doesn't appear to be probabilistic—the observations of classically chaotic orbits of Hyperion don't seem to correspond to probabilistic predictions of quantum theory—might such observations still have smuggled in claims about probabilities, and be relying on such claims in order to underwrite their empirical confirmation?

If there were, say, emergent branches on which Hyperion displayed some other behaviour from that predicted above, perhaps some would claim that the possibility of observations on such branches undermines the claimed empirical grounding of the amplitude-irrelevance link.

I should say, first, that I don't know of any good reason to suppose that there are any such branches. One would need to show that there is some decohered dynamical system, approximately screened off from the other branches, on which a stable

Hyperion exists and exhibits some other dynamics. While I cannot rule out such a possibility, I needn't take it especially seriously until a physical justification for doing so has been provided.

But even were there many branches on which Hyperion had lots of different behaviours, and so some interpretation of probabilities were required to generate expectations as to what an observer faced with a Hyperion splitting event will observe, my argument is not undermined.

What *we* observe is a classically chaotic Hyperion, and I've given reasons above to think that this emerges as a stable macroscopic system from the underlying quantum state. This is alone sufficient to establish that emergence may be non-circularly justified. That other possible observers may observe something different, and that, looking forward, we may be unsure what to expect absent an interpretation of probability, are both claims compatible with the emergence being empirically grounded.

The Born rule measure, as applied in contexts where a probabilistic interpretation of the mod-squared amplitudes are ruled out, is central to the derivation of the (very many) branches where Hyperion is classically chaotic as well as the hypothetical branches where Hyperion behaves some other way. All sorts of worries may follow about confirmation and uncertainty—issues which I do not have space to explore here—but the justification of emergence and the link between what we in fact observe and the conditional irrelevance of the small-amplitude contributions is borne out. It's justified by the role it plays in explaining what we observe!

Insofar as my aim is to undermine the putative vicious circularity, the bar is fairly low. The goal was to show that on reasonable assumptions we can determine the existence of emergent classical structure, and that no probabilities essentially feature in the derivation.

As noted, Dawid and Thébault's argument depends significantly on principle **Z**— the claim that the empirical contact of Everettian quantum mechanics must come via probabilities. It's worth noting that their objection to the Everettian story is particularly focused on a putative conflict between the decision-theoretic approach to probability and the decoherence account of emergence. Therefore it's significant that I developed the above with reference to empirical observations that are not probabilistic in nature. So, insofar as I am proposing that one can sidestep decision-theoretic justifications of probability in justifying emergence claims, I am not directly disagreeing with their main thesis. Perhaps we really disagree only to the extent that they suppose the decision-theoretic approach is the only possible justification for applications of the Born rule measure.

## 6. Emergence?

There are other critiques of the Everett interpretation of quantum mechanics that focus more squarely on its appeal to emergence rather than on the relation between emergence and probabilities. It's worth considering these because they are also, at least partially, defused by the approach to emergence set out above.

Maudlin (2010) develops the interesting worry that the Everettian approach is inadequate while it doesn't provide a primitive ontology—the suggestion is that in the absence of any additional stuff that can be arranged in, say, the configurations to which the wavefunction assigns amplitudes in configuration space, the theory lacks empirical content:

> any theory whose physical ontology is a complete wavefunction monism automatically inherits a severe interpretational problem: if all there is is the wavefunction, an extremely high-dimensional object evolving in some specified way, *how does that account for the low-dimensional world of objects that we start off believing in, whose apparent behaviour constitutes the explanandum of physics in the first place?* ... the obvious way for a physical theory to accomplish this task is to postulate that there *are* localized objects in a low-dimensional spacetime. (Maudlin 2010, 132–3, original emphasis)

In sections 2–4 I set out how, at least in one particular case, the evolution of the wavefunction can systematically give rise to approximately localised quasi-classical ontology. So the tools of emergence allow us to recover (approximately) localised objects without these having been postulated in addition. This analysis is the result of inspection as to the circumstances under which we may interpret parts of the complex, more fundamental ontology as localised and empirically grounded.

So, do we need to posit non-emergent localised ontology in addition to the wavefunction in order to have an ontology that may be interpreted as such and that can ground the empirical content of the theory? In my view the answer to this question depends on whether we are to view Everettian quantum theory as fundamental or emergent.

Were the theory fundamental then Maudlin is right that more ought to be said about the character of the ontology on offer. One might, for example, follow Wallace (2012, 315) in advocating structural realism, and something along these lines is discussed in Franklin and Robertson (2021); though even structural realism is compatible with different answers to the question "is it structure all the way down?"

But we have good, well-known reasons to think that even the quantum field theories of the standard model are not fundamental theories. As such, the right way to view Everettian quantum theory is as itself emergent. Wallace (2020) compellingly argues that quantum theory is best understood as a framework theory; thus, rather than seeking to characterise the ontology of the theory in general we should consider how each of its instantiations—for example, systems of qubits, non-relativistic particle mechanics, phonons, superfluids, quantum fields, ...—emerge from other quantum theories or, say, from some more fundamental theory of quantum gravity. On the account developed above, just in case we have novel dependencies that screen off lower-level details we ought to view the entities involved in those dependencies as real and emergent. So, for example, phonons emerge as entities when, say, the Boltzmann equation for phonon thermal transport is satisfied and atomic displacements are screened off.

Once one identifies the particular quantum theory under discussion by filling out the framework with details relevant to the particular domain of inquiry, the specific emergence relations to some more fundamental quantum or non-quantum theory may be determined. Attempting to provide an ontology for quantum theory simpliciter is a mistaken enterprise.

An interpretation of quantum mechanics, qua non-fundamental theory, certainly shouldn't be held to the standards of providing an interpretation of fundamental reality. Either it's structure/patterns all the way down, or some future, say, quantum gravity theory, will furnish the fundamental constituents of reality. Either way, Everettian quantum mechanics is in good standing.

We can consequently respond to Maudlin by pointing out that neither the underlying quantum theories nor the localised classical entities are fundamental, both are emergent and, as such, are empirically grounded by virtue of their experimental predictions at various distinct levels. The account presented and defended in the first five sections of this paper demonstrates a number of ways that empirical grounding is to be achieved: in particular, environment-induced decoherence leads to localisation and the emergence of classical behaviour, which is subsequently observed. Given a broad commitment to theoretical reductionism I think that it will be possible to relate each theory to underlying theories, though often in non-trivial ways. However, reduction is not required for grounding the empirical content of the theory. Since emergent theories are usually discovered and empirically substantiated in advance of their more fundamental counterparts, how could the empirical content of our theories require knowing what they are "about" at the fundamental level?

Like every other non-fundamental theory we've ever met (that is, every theory we've met) quantum theory emerges from some more fundamental theory. But the fact that biology is emergent from more fundamental physics and chemistry does not preclude biology's making empirical contact with reality nor its ontological interpretation. In fact, biology makes its contact through its predictions about what will happen to parts of the world at its level. Quantum mechanics makes its empirical contact in a similar way.

There are two related worries within Maudlin's approach. First, that the whole emergence framework presupposed here cannot work—the response that was just presented is to draw analogies with instances of emergence elsewhere in science and to argue that this case is in no way special. If this response is to be refuted, a new way of thinking about emergence across science must be offered. Indeed, Maudlin (2015, 356) notes that the problem of defining the local characteristics of non-microscopic systems is "easily and transparently solved by simple aggregation of microscopic parts."

However, as it stands this is not a fully fledged account of emergence and it's doubtful that it can account for the ontology of the special sciences; by contrast, the approach set out in section 2 is designed to accommodate the fact that macroscopic entities are defined by their dependencies and functional relationships rather than the parts of which they're constituted—in fact, no account of macroscopic entities merely in terms of their parts is likely to succeed, as interactions and dynamics play an essential role in any story of the emergence of higher-level properties such as hardness, temperature, pressure, etc. If Maudlin's goal is to replace all such accounts with one in terms of "simple aggregation" then his project certainly has a long way to go.

Maudlin's second worry is specific to quantum theory. Maudlin may draw the disanalogy with other sciences as follows: each really is about the fundamental description—for example, even though it's difficult to tell, the subject matter of biology is arrangements of fundamental stuff, and the empirical grounding of biology happens via such arrangements. We know, however, that Everettian quantum theory can't be quite like that, because we know that its subject matter isn't fundamentally local. Without positing Bohmian particles or some other localised primitive ontology, Maudlin suggests that we don't have any ontology sufficiently localised to be able to play the required role. But what reason is there to think that the fundamental stuff of quantum mechanics has to be localised?

All we need to show is that quantum mechanics is as localised as it's observed to be, and the decoherence-based account has the capacity to show this. Replying further to this second objection would take me beyond the scope of this paper, though see Ney and Phillips (2013) for further remarks on this matter. Suffice it to say that I think the emergence of effectively localised classical systems, as described above, is sufficient to show that no fundamental exactly localised matter is required. The model discussed in section 4 shows that entangled non-local systems can give rise to effectively localised systems. Since we don't have a fundamental theory, it may be that metaphysical analysis of quantum mechanics is premature, though this depends on the assumption—with which I disagree—that metaphysics exclusively concerns the fundamental. My goal here has been to provide a rival non-fundamental, emergent ontology that can make sense of how we have the array of quantum ontologies that we have, from fields to qubits and so on.

Overall, the contest is between rival frameworks. What I hope to have done in this paper is to show that the Everettian claims to emergence fit well within a much more general account of emergence in science. The competing approaches, such as that due to Maudlin, need far more work to evidence similar claims.

A related set of concerns are raised by Monton (2013, 164) in response, especially, to analysis in Wallace and Timpson (2010):

> It's simply not the case that one can have a 3$N$-dimensional space with a field evolving in it, such that when the field has a configuration, three-dimensional objects come into existence…for them to come into existence emergently, without this happening in accordance with certain novel laws of physics, is not the way a world where quantum mechanics is true works.

I don't wish to take a view on the dimensionality of the space in which the wavefunction or quantum state lives fundamentally, but I do take issue with Monton's assertion that three-dimensional objects cannot come into existence emergently without some novel laws of physics. Or, at least, there are two ways to interpret that claim: first, that it requires novel fundamental (strongly emergent) laws, and I claim that these are not required; or second (though I doubt this is Monton's intended reading), weakly emergent laws are required for the existence of three-dimensional objects. I agree with the latter interpretation, and claim that these are the classical mechanical laws with which we are familiar.

It's just not the case that we need special laws to tell us when there is emergent ontology—in fact, emergent ontology is vague and thus precise determinate laws specifying when this is to be found would be inappropriate. What we have in this paper is a prescription that tells us when we do and don't have local quasi-classical ontology. This corresponds to the emergence of ontology in other non-physical sciences, and I don't know what else we ought to be looking for. If Monton prefers only to accept fundamental ontology, then his ontology is currently empty! If, on the other hand, Monton accepts that there is non-fundamental ontology, then why shouldn't we grant that status to Hyperion, emerging as it does out of the more fundamental quantum description as a result of novelty and screening off?

## 7. Conclusion

Everettian quantum mechanics posits an extravagant ontology, and, of course, such an ontology requires significant empirical confirmation to be taken seriously. While it's often claimed that the empirical success of quantum theory is evidence for the Everett interpretation, the empirical success of the emergent classical dynamics can be viewed in a similar light. This helps defuse some of the objections raised against decoherence approaches to emergence. Some such objections presuppose that evidence for quantum theory only comes via probabilities. By focusing on the case of Hyperion, where the prediction of quantum mechanics together with decoherence is deterministic—that the orbit will be classically chaotic—we see that putative vicious circularities may be avoided.

Overall, the rebuttal of objections to the Everettian emergence story relied on the clarification of the sense in which the Everettian worlds emerge, as set out above. Through a focus on the fact that emergence is a consequence of screening off, and screening off is a relative phenomenon—the quantum corrections are screened off relative to the classical dynamics—the objections that we need some independent justification of the interpretation of small amplitudes as negligible is undermined. Rather, we discover that the amplitudes' magnitude is related to their dynamical salience and that this claim is evidenced by the outcomes of derivations and the observation of emergent classical behaviour.

By understanding the emergence of the Everettian multiverse along the lines of emergence in other scientific theories, I hope to have established that the claims of the theory may receive empirical confirmation. Even without an interpretation of probabilities in Everettian quantum mechanics, claims to emergence can be justified.

**Acknowledgements.** I'm very grateful to Eleanor Knox, Katie Robertson, and Karim Thébault, as well as audiences at the University of Bristol, UC Irvine, BSPS Conference, University of Bergen, University of Birmingham, and two anonymous referees for helpful feedback.

## References

Baker, David J. 2007. "Measurement Outcomes and Probability in Everettian Quantum Mechanics." *Studies In History and Philosophy of Science Part B: Studies In History and Philosophy of Modern Physics* 38(1):153–69. doi: 10.1016/j.shpsb.2006.05.003.

Barrett, Jeffrey A. 2011. "Everett's Pure Wave Mechanics and the Notion of Worlds." *European Journal for Philosophy of Science* 1(2):277–302. doi: 10.1007/s13194-011-0023-9.

Berry, Michael V. 2001. "Chaos and the Semiclassical Limit of Quantum Mechanics (Is the Moon There When Somebody Looks?)." In *Quantum Mechanics: Scientific Perspectives on Divine Action*, edited by Robert J. Russell, Philip Clayton, Kirk Wegter-McNelly, and John Polkinghorne, 41–54. Tucson, AZ: Vatican Observatory Publications.

Crull, Elise M. 2021. "Quantum Decoherence." In *The Routledge Companion to Philosophy of Physics*, edited by Eleanor Knox and Alastair Wilson. London: Routledge.

Dawid, Richard, and Karim P. Y. Thébault. 2015. "Many Worlds: Decoherent or Incoherent?" *Synthese* 192(5):1559–80. doi: 10.1007/s11229-014-0650-8.

Evans, Peter W., and Karim P. Y. Thébault. 2020. "On the Limits of Experimental Knowledge." *Philosophical Transactions of the Royal Society A* 378(2177):20190235. doi: 10.1098/rsta.2019.0235.

Franklin, Alexander and Eleanor Knox. 2018. "Emergence Without Limits: The Case of Phonons." *Studies In History and Philosophy of Modern Physics* 64:68–78. doi: 10.1016/j.shpsb.2018.06.001.

Franklin, Alexander, and Katie Robertson. 2021. "Emerging into the Rainforest: Emergence and Special Science Ontology." Preprint, available at http://philsci-archive.pitt.edu/19912/.

Franklin, Alexander, and Vanessa Seifert. 2020. "The Problem of Molecular Structure Just Is the Measurement Problem." *The British Journal for the Philosophy of Science.* doi: 10.1086/715148.

Habib, Salman, Kosuke Shizume, and Wojciech H. Zurek. 1998. "Decoherence, Chaos, and the Correspondence Principle." *Physical Review Letters* 80(20):4361. doi: 10.1103/PhysRevLett.80.4361.

Halliwell, Jonathan J. 1998. "Decoherent Histories and Hydrodynamic Equations." *Physical Review D* 58(10):105015. doi: 10.1103/PhysRevD.58.105015.

Halliwell, Jonathan J. 2010. "Macroscopic Superpositions, Decoherent Histories, and the Emergence of Hydrodynamic Behaviour." In Saunders et al. 2010, 99–117.

Joos, Eric, H. Dieter Zeh, Claus Kiefer, Domenico J. Giulini, Joachim Kupsch, and Ion-Olimpiu Stamatescu. 2013. *Decoherence and the Appearance of a Classical World in Quantum Theory.* New York: Springer.

Kent, Adrian. 2010. "One World Versus Many: The Inadequacy of Everettian Accounts of Evolution, Probability, and Scientific Confirmation." In Saunders et al. 2010, 307–54.

Knox, Eleanor. 2016. "Abstraction and its Limits: Finding Space for Novel Explanation." *Noûs* 50(1):41–60. doi: 10.1111/nous.12120.

Ladyman, James, Don Ross, David Spurrett, and John Collier. 2007. *Every Thing Must Go: Metaphysics Naturalized.* Oxford: Oxford University Press.

Lin, W. and Leslie Ballentine. 1990. "Quantum Tunneling and Chaos in a Driven Anharmonic Oscillator." *Physical Review Letters* 65(24):2927.

Maudlin, Tim. 2010. "Can the World Be Only Wavefunction?" In Saunders et al. 2010, 121–43.

Maudlin, Tim. 2015. "The Universal and the Local in Quantum Theory." *Topoi* 34(2):349–58. doi: 10.1007/s11245-015-9301-z.

Monton, Bradley. 2013. "Against 3n-Dimensional Space." In *The Wave Function: Essays on the Metaphysics of Quantum Mechanics*, edited by Alyssa Ney and David Z. Albert. Oxford: Oxford University Press.

Ney, Alyssa, and Kathryn Phillips. 2013. "Does an Adequate Physical Theory Demand a Primitive Ontology?" *Philosophy of Science* 80(3):454–74. doi: 10.1086/671076.

Ross, Don. 2000. "Rainforest Realism: A Dennettian Theory of Existence." In *Dennett's Philosophy: A Comprehensive Assessment*, edited by Don Ross, Andrew Brook, and David L. Thompson. Cambridge, MA: MIT Press.

Saunders, Simon. 2021. "The Everett Interpretation: Structure." In *The Routledge Companion to Philosophy of Physics*, edited by Eleanor Knox and Alastair Wilson. London: Routledge.

Saunders, Simon, Jonathan Barrett, Adrian Kent, and David Wallace (eds). 2010. *Many Worlds? Everett, Quantum Theory, and Reality.* Oxford: Oxford University Press.

Schlosshauer, Maximilian. 2008. "Classicality, the Ensemble Interpretation, and Decoherence: Resolving the Hyperion Dispute." *Foundations of Physics* 38(9):796–803. doi: 10.1007/s10701-008-9237-x.

Schlosshauer, Maximilian. 2019. "Quantum Decoherence." *Physics Reports* 831:1–57. doi: 10.1016/j.physrep.2019.10.001.

Wallace, David. 2012. *The Emergent Multiverse: Quantum Theory According to the Everett Interpretation.* Oxford: Oxford University Press.

Wallace, David. 2013. "A Prolegomenon to the Ontology of the Everett Interpretation." In *The Wave Function: Essays on the Metaphysics of Quantum Mechanics*, edited by Alyssa Ney and David Z. Albert. Oxford: Oxford University Press.

Wallace, David. 2014. "Probability in Physics: Statistical, Stochastic, Quantum." In *Chance and Temporal Asymmetry*, edited by Alastair Wilson. Oxford: Oxford University Press.

Wallace, David. 2018. "Spontaneous Symmetry Breaking in Finite Quantum Systems: A Decoherent-Histories Approach." Available at https://philsci-archive.pitt.edu/14983/1/ssbfinite.pdf.

Wallace, David. 2020. "Lessons from Realistic Physics for the Metaphysics of Quantum Theory." *Synthese* 197(10):4303–18. doi: 10.1007/s11229-018-1706-y.

Philosophy of Science    309

Wallace, David. 2022. "Stating Structural Realism: Mathematics-First Approaches to Physics and Metaphysics." *Philosophical Perspectives* 36(1):345–78. doi: https://doi.org/10.1111/phpe.12172.

Wallace, David, and Christopher G. Timpson. 2010. "Quantum Mechanics on Spacetime I: Spacetime State Realism." *The British Journal for the Philosophy of Science* 61(4):697–727. doi: 10.1093/bjps/axq010.

Wiebe, Nathan, and Leslie Ballentine. 2005. "Quantum Mechanics of Hyperion." *Physical Review A* 72(2):022109. doi: 10.1103/PhysRevLett.65.2927.

Wilson, Mark. 2006. *Wandering Significance: An Essay on Conceptual Behaviour.* Oxford: Oxford University Press.

Wimsatt, William C. 2007. *Re-Engineering Philosophy for Limited Beings: Piecewise Approximations to Reality.* Cambridge, MA: Harvard University Press.

Woodward, James. 2021. "Explanatory Autonomy: The Role of Proportionality, Stability, and Conditional Irrelevance." *Synthese* 198(1):237–65.

Zurek, Wojciech H. 1998. "Decoherence, Chaos, Quantum-Classical Correspondence, and the Algorithmic Arrow of Time." *Physica Scripta* 1998(T76):186. doi: 10.1238/Physica.Topical.076a00186.

Zurek, Wojciech H. 2003. "Decoherence, Einselection, and the Quantum Origins of the Classical." *Reviews of Modern Physics* 75:715–75. 10.1103/RevModPhys.75.715.

Zurek, Wojciech H. and Juan Pablo Paz. 1994. "Decoherence, Chaos, and the Second Law." *Physical Review Letters* 72(16):2508. doi: 10.1103/PhysRevLett.72.2508.